\documentclass{article}
\usepackage{amsmath,amsthm}
\usepackage{amssymb,latexsym}
\usepackage[mathscr]{eucal}
\usepackage{setspace}
\usepackage{graphics}
\usepackage{array}

\newcommand{\pa}{\partial}

\newcommand{\na}{\nabla}

\newcommand{\lp}{\left(}
\newcommand{\rp}{\right)}

\newcommand{\beq}{\begin{equation}}
\newcommand{\eq}{\end{equation}}
\newcommand{\bs}{\boldsymbol}
\newcommand{\ti}{\times}

\begin{document}

\title{\textbf{Kelvin's Canonical Circulation Theorem in Hall Magnetohydrodynamics}}         
\author{B. K. Shivamoggi\\
University of Central Florida\\
Orlando, FL 32816-1364, USA\\
}        
\date{}          
\maketitle

\noindent \large{\bf Abstract} \\ \\

The purpose of this paper is to show that, thanks to the restoration of the legitimate connection between the current density and the plasma flow velocity in Hall magnetohydrodynamics (MHD), Kelvin's Circulation Theorem becomes valid in Hall MHD. The ion-flow velocity in the usual circulation integral is now replaced by the canonical ion-flow velocity.

\pagebreak

\noindent\large\textbf{1. Introduction}\\

The magnetohydrodynamics (MHD) model (Cowling \cite{Cow}) offers the simplest framework by which the macroscopic interactions between plasmas and magnetic fields may be explored. Though it is not a very good approximation to any real plasma, many plasma phenomena are indeed well described  by a fluid model. One awkward issue of basic conceptual nature with the MHD model is the assumption that the current density is only connected with the magnetic field (via Ampere's law) but somehow not with the plasma velocity. This anomaly appears to be the cause of a very complicated Haniltonian analytic structure for the MHD model (the Poisson bracket, in particular, Shivamoggi \cite{Shi}). Another example is Kelvin's Circulation Theorem (Batchelor \cite{Bat}) which does not hold unless the closed material curve (around which the flow circulation is calculated) is a closed magnetic field line (Shercliff \cite{She}).

On the other hand, in the MHD model, ions dominate the dynamics, while electrons merely serve to shield out rapidly any charge imbalances. However, in a high-$\beta$ plasma, on length scales in the range $d_e<\ell<d_i$, where $d_s$ is the skin depth, $d_s\equiv c/\omega_{ps}$, ($s=i,e$ referring to the ions and electrons, respectively) the electrons decouple from the ions. This leads to an additional transport mechanism for the magnetic field via the Hall current (Sonnerup \cite{Son}), which is the ion inertia contribution in Ohm's law. The Hall effect leads to the generation of whistler waves whose,
\begin{itemize}
  \item [*] frequency lies between ion-cyclotron and electron-cyclotron frequencies $\omega_{c_i}$ and $\omega_{c_e}$, respectively,
  \item [*] phase velocity exceeds that of Alfv\'{e}n waves for wavelengths parallel to the applied magnetic fields less than $d_i$.\end{itemize} 

Further, the decoupling of ions and electrons in a narrow region around the magnetic neutral point (where the ions become demagnetized while the electrons remain magnetized) allows for a rapid electron flows in the ion-dissipation region and hence a faster magnetic reconnection process in the Hall MHD regime (Mandt et al. \cite{Man}).

The purpose of this paper is to show that the Hall MHD model supports Kelvin's Circulation Theorem without placing any restriction on the closed material curve, provided the ion-flow velocity is replaced by the canonical ion-flow velocity in the circulation integral\footnote{Similar result was shown to hold for the electron MHD (EMHD) model (Mahajan and Yoshida \cite{Mayo}) which is applicable to length scales $\rho_e\ll\ell\ll \rho_i,$ $\rho_s$ being the cyclotron radius, where electrons dominate the dynamics while demagnetized ions merely serve to provide the neutralizing static background (Kingsep et al. \cite{Kin}, Gordeev et al. \cite{Gor}).}.

\vspace{.20in}

\noindent\large\textbf{2. Kelvin's Circulation Theorem in Hall MHD}\\

The Hall MHD equations (which were originally formulated by Lighthill \cite{Lig} following his far-sighted recognition of the importance of the Hall term in the generalized Ohm's law) are (in usual notation), 

\beq
m_i n\left[\frac{\pa{\bf v}_i}{\pa t}+\lp {\bf v}_i\cdot\na\rp {\bf v}_i\right]=-\na\lp p_i+p_e\rp+\frac{1}{c}\phantom{x}{\bf J}\times {\bf{B}}
\eq

\beq
{\bf E}+ \frac{1}{c}\phantom{x}{\bf v}_i\times{\bf{B}}=\frac{1}{nec}\phantom{x}{\bf J}\times {\bf{B}}
\eq

Equations (1) and (2) admit a Hamiltonian formulation (Shivamoggi \cite{Shi}) and an impulse formulation (Shivamoggi \cite{Shi2}).

Using equation (2), equation (1) may be rewritten as

\beq
m_i n \frac{d{\bf v}_i}{d t}=-\nabla\lp p_i+p_e\rp+ne\lp{\bf E}+\frac{1}{c}\phantom{x}{\bf v}_i\phantom{x}\times {\bf{B}}\rp
\eq

\noindent or

\beq
\frac{d{\bf V}_i}{d t}=-\lp \frac{p_i+p_e}{m_i n}\rp+\frac{e}{m_i c}{\bf v}_i\times{\bf B}+\frac{e}{m_i n}\lp {\bf v}_i\cdot\nabla\rp {\bf A}
\eq

\noindent where, ${\bf V}_i$ is the canonical ion-flow velocity,

\beq
{\bf V}_i\equiv {\bf v}_i+\frac{e}{m_i c}\phantom{x} {\bf A}
\eq
	
\noindent and

\beq
{\bf E}=-\frac{1}{c}\frac{\pa{\bf A}}{\pa t},\phantom{x}\frac{d}{dt}\equiv\frac{\pa}{\pa t}+\lp{\bf v}_i\cdot\nabla\rp
\eq

\noindent and we have assumed the flow to be barotropic.

Consider the canonical circulation for Hall MHD defined by

\beq
\Gamma\equiv\underset{C}\oint{\bf V}_i\cdot d\ell
\eq
	
\noindent where $C$ is a closed curve made up of the same ions, and is so convected with the ion-flow velocity ${\bf v}_i$.

We then have, 

\beq
\frac{d\Gamma}{dt}=\underset{C}\oint\phantom{x}\left[\frac{d{\bf V}_i}{dt}\cdot d{\bs{\ell}}+{\bf V}_i\cdot\lp d{\bs{\ell}}\cdot\nabla\rp{\bf v}_i\right]
\eq

\noindent Using equation (4), (8) becomes

\beq
\begin{matrix}
\displaystyle\frac{d\Gamma}{dt}=\underset{C}\oint d{\bs{ \ell}}\cdot\left[-\nabla\lp\frac{p_i+p_e}{m_i n}\rp+\frac{e}{m_i c}{\bf v}_i\times{\bf B}\right.\\
\\
\displaystyle\left.+\frac{e}{m_i c}\lp {\bf v}_i\cdot\nabla\rp {\bf A}\right]+\underset{C}\oint {\bf V}_i\cdot\lp d{\bs{\ell}}\cdot\nabla\rp {\bf v}_i
\end{matrix}
\eq

\indent Noting, 

\beq
\begin{matrix}
\displaystyle{\bf V}_i\cdot\lp d{\bs {\ell}}\cdot\nabla\rp {\bf v}_i=-{\bf V}_i\cdot\left[d{\bs{\ell}}\times\lp\nabla\times{\bf v}_i\rp\right]+d{\bs{\ell}}\cdot\lp{\bf V}_i\cdot\nabla\rp {\bf v}_i\\
\\
\displaystyle=d{\bs{\ell}}\cdot\left[{\bf V}_i\times{\bs \omega}_i+\lp{\bf V}_i\cdot\nabla\rp{\bf v}_i\right]
\end{matrix}
\eq

\noindent (9) becomes, 

\beq
\begin{matrix}
\displaystyle\frac{d\Gamma}{dt}=\underset{C}\oint d{\bs{ \ell}}\cdot\left[-\nabla\lp\frac{p_i+p_e}{m_i n}\rp\frac{e}{m_i c}{\bf v}_i\times{\bf B}\right.\\
\\
\displaystyle+\frac{e}{m_i c}\left\{\lp {\bf v}_i\cdot\nabla\rp {\bf A}+\lp {\bf A}\cdot\nabla\rp{\bf v}_i\right\}+\left.\frac{e}{m_i c}{\bf A}\times{\bs\omega}_i+\nabla\lp\frac{{\bf v}^2_i}{2}\rp\right]
\end{matrix}
\eq

\noindent where ${\bs\omega}_i$ is the ion-flow vorticity, 

\beq
{\bs\omega}_i\equiv\na\ti{\bf v}_i.
\eq

\indent (11) may be simplified to give

\beq
\begin{matrix}
\displaystyle\frac{d\Gamma}{dt}=\underset{C}\oint d{\bs{\ell}}\cdot\nabla\left[\frac{{\bf v^2_i}}{2}+\frac{e}{m_i c}\lp {\bf v}_i\cdot {\bf A}\rp-\left(\frac{p_i+p_e}{m_i n}\right)\right]=0.
\end{matrix}
\eq
	
\noindent(13) implies that the canonical circulation around any closed curve made up of the same ions is an invariant in Hall MHD. It is of interest to note that the quantity in the bracket in the integral in (13) represents the Lagrangian for the plasma particles (with the neglect of electron inertia).

\vspace{.20in}

\noindent\large\textbf{4. Discussion}\\
	
Thanks to the disconnect between the current density and the plasma flow velocity, the MHD model ends up possessing a very complicated Hamiltonian analytic structure. Another setback for the MHD model is the failure of Kelvin's Circulation Theorem to hold unless the closed material curve in question is a closed magnetic field line (Shercliff \cite{She}). Restoration of the legitimate connection between the current density and the plasma flow velocity as in Hall MHD in the present paper (and EMHD, Mahajan and Yoshida \cite{Mayo}) leads to simplification of the Hamiltonian analytic structure (Shivamoggi \cite{Shi2}) and validity of Kelvin's Circulation Theorem provided the canonical ion-flow velocity now appears in place of ion-flow velocity in the circulation integral.

\vspace{.20in}
\newpage
\noindent\large\textbf{Acknowledgements}\\

I am thankful to Professor S. M. Mahajan for several stimulating discussions on this problem.

\end{document}